\documentclass{llncs}

\usepackage{listings}
\usepackage{courier}
\usepackage{graphicx}
\usepackage{verbatim}
\usepackage{dsfont}

\lstdefinelanguage{scala}{
  morekeywords={abstract,case,catch,class,def,%
    do,else,extends,false,final,finally,%
    for,if,implicit,import,match,mixin,%
    new,null,object,override,package,%
    private,protected,requires,return,sealed,%
    super,this,throw,trait,true,try,%
    type,val,var,while,with,yield},
  otherkeywords={=>,<-,<\%,<:,>:,\#,@},
  sensitive=true,
  morecomment=[l]{//},
  morecomment=[n]{/*}{*/},
  morestring=[b]",
  morestring=[b]',
  morestring=[b]"""
}
\begin{document}
\newcommand{\code}[1]{\textsf{#1}}

\pagestyle{headings}

\title{An Open Framework for Extensible Multi-Stage Bioinformatics Software}
\author{Gabriel Keeble-Gagn\`ere\inst{1} \and Johan Nystr\"om-Persson\inst{2}
\and \\ Matthew Bellgard\inst{1} \and Kenji Mizuguchi\inst{2}}

\institute{Centre for Comparative Genomics, Murdoch University, Australia
\and
National Institute of Biomedical Innovation, Japan}

\maketitle

\begin{abstract}
In research labs, there is often a need to customise software at every step in
a given bioinformatics workflow, but traditionally it has been difficult to
obtain both a high degree of customisability and good performance.
Performance-sensitive tools are often highly monolithic, which can make
research difficult. We present a novel set of software development principles
and a bioinformatics framework, Friedrich, which is currently in early
development. Friedrich applications support both early stage experimentation
and late stage batch processing, since they simultaneously allow for good
performance and a high degree of flexibility and customisability. These
benefits are obtained in large part by basing Friedrich on the multiparadigm
programming language Scala.  We present a case study in the form of a basic
genome assembler and its extension with new functionality. Our
architecture\footnote{Available freely under a dual GPL/MIT open-source license
from \url{https://bitbucket.org/jtnystrom/friedrich/}.} has the potential to
greatly increase the overall productivity of software developers and
researchers in bioinformatics.

\end{abstract}

\section{Introduction}

Bioinformatics poses a particularly difficult challenge for software
developers, with constantly changing end-user requirements and the
need to interact with an ever-expanding range of tools and data
formats. The advent of big data means that the tools and skills
required for data manipulation and basic research are now more
advanced than before. However, researchers are fundamentally
biologists and more interested in the data itself than in addressing
technical issues, which traditionally fall into the computer science
field. The challenge for software developers is thus to put the
maximum amount of power and flexibility in the hands of the users
while assuming as little technical knowledge as possible.

When large data volumes are processed, high performance software tools are
often used. However, such tools are often highly specialised and optimised for
a specific purpose, permitting only limited customisation. This kind of
software is often also \emph{monolithic}. Monolithic tools can be efficient for
handling big data problems, but such a design often runs counter to a natural
research process, since researchers often need to make adjustments to various
parts of the tools that they work with, particularly in fast-changing fields
such as bioinformatics. MacLean and Kamoun~\cite{MacLean2012}, reporting on
their experience bringing a small bioinformatics laboratory into the age of big
data, state that biologists at first tend to regard bioinformatics processes as
being monolithic, but once they understand their inner workings generally
become more productive, especially if they can take charge of tools and methods
themselves to some degree. Clearly, transparent and flexible tools have the
potential to play a very important role.

We argue that it is possible to develop software that makes
researchers more productive and enables them to ask more questions
about their data and their process by adopting a new set of software
development principles. In the following, we present the Friedrich
architecture (Section~\ref{sec:principles}).  We then discuss the
Friedrich framework, a toolkit for building bioinformatics
applications according to these principles
(Section~\ref{sec:implementation}).  We discuss the implementation of
a basic genome assembler based on Friedrich in
Section~\ref{sec:assembler}. We compare with other tools and
frameworks in Section~\ref{sec:comparison}, and conclude the paper
in Section~\ref{sec:conclusion}.

\section{The Friedrich software principles}
\label{sec:principles}

The Friedrich architecture is a set of interlocking software design
principles that, in our view, can support bioinformatics research very
effectively.


\textbf{Expose internal structure.} Bioinformatics software should
expose its internal building blocks and data flow to a high degree,
permitting reconfigurability.  Bioinformatics computation often
consists of sending data through a number of processing stages until
the desired output is produced. Frameworks should reflect this by
consisting of modules that can easily be rewired - reconnected in
different sequences - to represent changing workflows. This is the
opposite of a monolithic application, which is effectively a black
box.

\textbf{Conserve dimensionality maximally. } The processing of a given
data set~-- which can essentially be viewed as a set of points in a
mathematical space~-- to produce a given output, is analogous to a
\emph{projection} in geometry. For example, in $\mathds{R}^{3}$, the
equation

$$x_{1}^{2} + x_{2}^{2} + x_{3}^{2} = 1,$$

defines a sphere of radius 1 centred at the origin. The projection $proj_{1}$,
which sends $(x_{1}, x_{2}, x_{3}) \in \mathds{R}^{3}$ to $x_{1} \in
\mathds{R}$, when applied to the sphere defined above, yields: $x_{1}^{2} = 1,$
which defines the set of two points $\{ -1, 1 \}$. If $f: \mathds{R}^{3}
\rightarrow \mathds{R}^{3}$ is a mapping, then given a surface in
$\mathds{R}^{3}$ (such as the sphere defined above), the function $proj_{1}
\circ f$ returns an answer to the query ``At what points does the mapped
surface intersect the $x_{1}$-axis?'' Given an answer to the query, we cannot
extract information about the original surface.  In an analogous way, raw
bioinformatics data contains all possible information from a given experiment.
Thus it has \emph{maximum dimensionality}. As various data processing is
performed on this data set, its dimensionality is reduced.  For example, given
a set of reads from a DNA sequencing run, one processing step might be to
remove duplicates, to produce a set of non-redundant sequence reads.  This
would clearly reduce the dimensionality of the resultant data set, since the
redundancy information is lost. 


Maximal conservation of dimensionality permits users who are applying tools
experimentally to go back to previous stages of their computation and
attempt different parameters, adding a great deal of flexibility to
the experimental process, allowing new questions to be asked, and
saving time. It can also be thought of as
maximal preservation of the results of intermediate phases in the
computation. 


\textbf{Multi-stage applications.} Many tools need to be used in at
least two different stages, which may loosely be called
\emph{experimentation} and \emph{production}. In the experimental
stage, researchers explore newly available data in order to develop
methods and a basic understanding of what can be done. It is in this
stage that the need for customisation and flexibility is greatest. In
the production stage, a repeatable process is extracted and applied
systematically a large number of times. In this stage there is less
need for flexibility; instead, robustness, reliability, and
performance are valued.  However, \emph{a given analysis or tool, once
developed, often has to move across this boundary from the
experimental stage to the production stage}. This transition is often
nontrivial given that hitherto, incompatible technologies have often
been used in the two stages. In such a situation, one may opt to use
experimental stage technologies in both stages, resulting in poor
performance.  Alternatively, one may use production-stage technologies
in both stages, resulting in difficulty of experimentation. Finally,
one may re-develop the analysis from scratch once it makes the
transition, which would be a large additional effort.


Friedrich software should support a full range of development stages,
including experimentation, production, and any intermediate
points. Because a single technology framework is used consistently, it
becomes easy to move from experimentation to production, and also to
move back again. This enables a feedback loop between experimental
usage and production usage: when something unexpected occurs in the
large scale application of a tool, it can easily be taken back to the
workbench for inspection, and any adjustments made can be propagated
back again. Table~\ref{fig:comparison} gives a comparison.

\begin{table}
\centering
\footnotesize
\begin{tabular}{|p{2.6cm}|p{1.8cm}|p{2.6cm}|p{2.5cm}|p{2cm}|}
\hline
Context & Necessary flexibility & Typical programming language & Performance & Examples \\
\hline
\hline
Experimental stage tools & High & Perl, Python, R, ... & Low/ moderate & BioPerl, BioPython \\
\hline
Production stage tools & Low &  C, C++, Java, ... & Very high & Velvet, Abyss, BioJava \\
\hline
\emph{Friedrich} & High & Scala, Java & High & Section~\ref{sec:assembler} \\
\hline
\end{tabular}

\caption{A comparison of Friedrich's target characteristics with tools 
designed mainly for either experimentation or production.}
\label{fig:comparison}
\end{table}

\textbf{Flexibility with performance.} This is closely related to the
previous principle. If programming languages have traditionally been separable
into on one hand a category of high-performing but inflexible ones (in that
applications written in them are relatively hard to customise) and on the other
a category of poorly performing but flexible ones, we believe that the
relatively recent language Scala (see Section~\ref{sec:scala}) is an outlier
that provides for both good performance and high flexibility. This enables
flexibility with performance. For many bioinformatics applications, one should
not seek extreme performance or extreme flexibility but good levels of both.

\textbf{Minimal finality.} Monolithic software often makes
unsustainable assumptions about data formats, algorithm parameters and data
sizes. For example, the so-called next generation of sequencing equipment is
expected to render many of the current genome analysis software tools
unusuable, largely for the reason that certain quantity and size parameters
will change. Friedrich applications should assume a minimum of finality.
Software developers should not dictate how the framework or its building blocks
should ultimately be used, since they cannot possibly anticipate all the usage
scenarios that may eventually appear. MacLean and Kamoun found that reorienting
research from a top-down model to a bottom-up model helped increase
productivity in the Sainsbury Laboratory~\cite{MacLean2012}. Minimising
finality also helps achieve this end.

\textbf{Ease of use.} Friedrich applications should not be hard for novices to
use. They should provide sensible defaults at all times, so that new
users can deploy them in common use cases with little effort.
Simplicity should not be sacrificed to the other principles.

We have now described the software design principles of the Friedrich
architecture. Next, we describe our implementation of the Friedrich
framework, as well as an application built on top of it.

\section{The Friedrich framework}
\label{sec:implementation}
The Friedrich framework is implemented in the form of a Scala library
that permits users to develop bioinformatics applications easily. In
implementing this framework, our aim has been to allow application
developers to follow the principles we outlined in the previous
section easily. The framework is still under development, and this
section describes its current state. 

\subsection{The Scala programming language}\label{sec:scala}
An early decision was made to base Friedrich on Scala, a novel programming
language for the Java virtual machine, which is being developed by Martin
Odersky and others~\cite{scala} (\url{http://www.scala-lang.org}). Programming
languages are traditionally classified as \emph{functional} or
\emph{imperative}. Functional languages emphasise avoidance of side effects and
composition of functions. Imperative languages, such as Perl, C, and Java, have
been more widely used in the mainstream, and generally functions in these
languages may have side effects. Scala blends these two paradigms. It provides
libraries, constructs and idioms for stateless, purely functional programming
as well as for stateful, imperative, object-oriented programming. Scala code is
often very compact compared with equivalent Java code, and, provided that the
programmer is somewhat disciplined, can be highly readable. 

Scala brings several important benefits to Friedrich.
\begin{itemize}
\item Scala provides for high programmer productivity and is very well suited
    to big data tasks, performing well~\cite{37122} even under heavy loads,
    thanks to the maturity of the underlying Java platform.
\item Existing Java libraries for tasks such as graph processing, database
    access, calculation and so on can be taken advantage of immediately.
\item Because of its strong support for functional programming and immutable
    state, Scala is a foundation that lends itself well to parallel processing,
    the need for which cannot be ignored in bioinformatics today.
\end{itemize}

Scala has much of the flexibility and productivity of scripting
languages such as Ruby, Python and Perl. For example, Scala has
features such as an interactive interpreter with auto-completion,
pattern matching and convenient regular expression support. Type
inference means that types in many cases do not need to be declared.
SBT (Simple Build Tool), which is widely used by the Scala community, permits
automatic dependency management and library downloading in a style
that resembles Perl's well-known CPAN package repository.


In a survey of software engineering techniques used in 22 different
bioinformatics software projects, Rother et al. described 12 practices
that were found to be useful~\cite{citeulike:9595647}. Scala and
Friedrich directly support many of these, benefiting both from the
mature development tools available for the Java platform and from its
own tools. For example, Scala has good support for unit testing and a
sophisticated documentation generator, and Friedrich supports
practices such as frequent release and feedback cycles, since it
enables easy transitions between the experimental and production
stages.

\subsection{Friedrich application components}
Friedrich contains the following key components for building applications.

\begin{description}
\item[Phases and pipelines.] Friedrich applications are organised as
sets of \emph{phases}, according to the model illustrated in
Figure~\ref{fig:comparison}. Sequences of phases are called
\emph{pipelines}. Friedrich provides foundational classes that can be
extended to implement new phases, as well as functions for managing
and running pipelines.
\item[Data object classes.] Friedrich phases operate on standardised
data objects. For a given application, all experimental data as well
as configuration parameters is stored in these objects.
\item[Configuration management.] Pipelines and general application
parameters are stored in XML configuration files
(Figure~\ref{fig:pipeline}).  Friedrich provides facilities for reading
these configurations and automatically creating pipelines from them.
\item[Core bioinformatics functionality.] Friedrich provides a small
library of core bioinformatics algorithms and data representations.
\end{description}

In order to implement a new Friedrich application, one should select a
data object type or define a new one, implement the necessary phases,
and write a \code{main} method that invokes a pipeline using the
Friedrich API. As we will see, implementing phases is not difficult.

Phases receive input that they make certain assumptions about
(\emph{phase preconditions}), perform some computation on it, and then
pass on this data in a new state (\emph{phase postconditions}) as
output. For example, our genome assembler makes use of phases such as
ScanReads, BuildGraph and FindPaths, among others (shown in
Figure~\ref{fig:phaseTree}). Phases can perform almost any
functionality. In accordance with our dimensionality principle, phases
should add information to the shared data object rather than remove or
overwrite. This permits the user to explore and manipulate the data
(in interactive mode) in between pipeline phases. Friedrich
applications can easily invoke pipelines based on their names only,
which means that workflows can be changed without recompiling an
application.

\lstset{language=XML}
\begin{figure}
\begin{lstlisting}
<settings>
  <pipeline name="default">
    <phase>miniasm.ScanReadsPhase</phase>		
    <phase>miniasm.BuildGraphPhase</phase>		
    <phase>miniasm.FindTipsPhase</phase>
    <phase>miniasm.ComputeCoveragePhase</phase>			
    <phase>miniasm.FindPathsPhase</phase>			
  </pipeline>
</settings>
\end{lstlisting}
\caption{An example of a pipeline configuration. The phases will be run in the
order shown. 'Miniasm' is the package name of the corresponding classes.}
\label{fig:pipeline}
\end{figure}

The components we have described support the six principles as much as
possible. Phases and pipelines are a natural way to expose structure.
When an application is made up of a set of relatively independent
phases, it becomes clear what its internal parts are, and the
configuration system permits them to be rewired easily. Conservation
of dimensionality is not enforced by the framework itself.  Phase
implementors are recommended to always add data to the shared data
object and not overwrite or remove it unless necessary.  In the
future, we plan to provide automatic data management facilities to
assist interactive use. Multi-stage applications and flexibility with
performance are benefits that we derive largely from our use of the
Scala language, as outlined above. Minimal finality is something we
obtain in part from Scala, and in part from the pipeline and phase
system, since the overall data flow of an application can be changed
at a late stage. Ease of use is a principle to be upheld by
application developers. 

\section{Genome assembly with Friedrich}
\label{sec:assembler}

\begin{figure}
\centering
\includegraphics[scale=1.1]{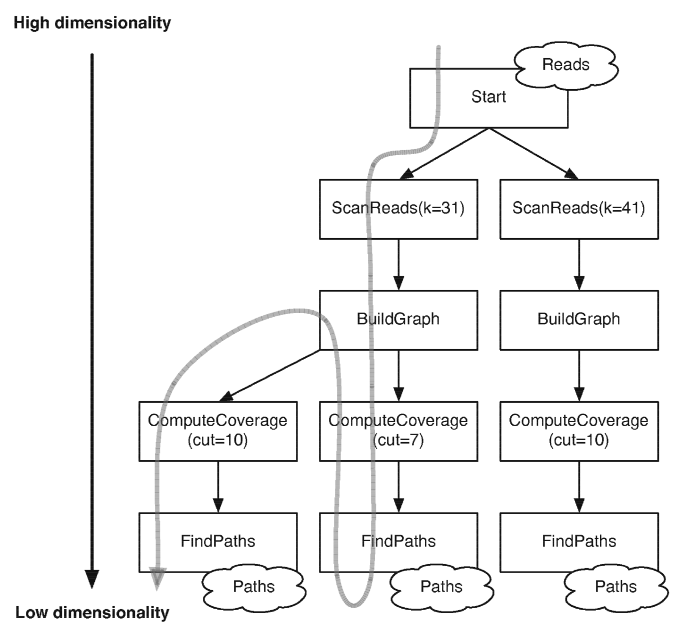}
\caption{The internal data flow of a hypothetical genome
assembler. The grey path illustrates how a user may wish to try a number of
different coverage cutoff values, which involves returning to a previous phase.}
\label{fig:phaseTree}
\end{figure}

Genome assembly refers to the process of turning raw sequence reads -- produced
from a sequencing run -- into contiguous regions of DNA, know as contigs, that
represent the original genome being analysed. In particular, \emph{de novo}
genome assembly refers to assembling a novel genome for the first time directly
from individual reads -- that is, without a reference genome to guide it.
Assembly methods have evolved from the \emph{overlap-layout-consensus} (OLC)
method (employed by early sequencing efforts, including the Human Genome
Project, which took advantage of the long reads produced by traditional Sanger
technology) to the \emph{de Bruijn} graph methods employed by most assemblers
that accept current high-throughput short read data. For the technical details
of genome assembly, we refer the reader to \cite{citeulike:10005303}. In short,
the nodes of the de Bruijn graph are sequences of length $k$ base pairs (known
as $k$-mers); an edge exists between two $k$-mers if their sequence overlaps by
$k - 1$ bases. This graph is then processed and contigs read off directly as
non-ambiguous paths.

One of the early motivations for Friedrich was the desire to investigate in
detail the inner workings of this process. Investigating assemblies with
commonly used assemblers such as Velvet~\cite{zerbino2008} and
ABySS~\cite{citeulike:4117809}, we found that output can vary considerably
given the same input data. As well as this, we found that outputs could vary
even on very small toy data sets (data not shown). Indeed, anyone who has used
these tools will be aware that different assemblers produce different output,
but rarely will the user have a clear idea of what exactly has been done
differently.

Figure~\ref{fig:phaseTree} outlines a simplified typical workflow for an
assembler. Internally, data is sent through a number of phases in order to
produce the final output.  As a rule, the output of each phase is less complex
than its input, and the final output is much simpler than the initial input.
This can be understood as a successive reduction of the dimensionality of the
data. Each phase within a tool such as this assembler can be controlled by
parameters (for example $k$, $cut$), and modifying the parameters of a phase
might affect the final output significantly. Thus, researchers might want to
traverse what we might call a \emph{phase tree} following the curved arrow in
order to compare outputs resulting from various configurations. In a monolithic
tool, this is generally not possible, since one cannot return to earlier phases
in the pipeline: the tool must be re-run from the starting point even when only
parameters of late phases are changed, if they can be changed at all. With
Friedrich, it is possible to interrogate the assembly at every step of the way. 

The Friedrich-based assembler that we have developed consists of an efficient
representation of sequences and reads, 11 processing phases and various utility
classes. The source code is about 3000 lines in length.

\subsection{Interactive use}
\label{sec:interactive}
The following is an example of an interactive Friedrich session to
process Illumina short read data\footnote{NCBI SRA experiment
ERX005938, run ERR015569. Only 1/9 of the reads were passed to
Friedrich.}. We launch the interactive Friedrich console using SBT.
If the source code of any phases or libraries being used has 
changed when Friedrich is launched
in this way, they will automatically be recompiled, permitting a
smooth development and testing workflow. The interactive Scala
environment has features such as tab-completion to show all available
alternatives. This environment evaluates Scala expressions as they are
typed in, and allows for functions and classes to be defined on the
fly.

\begin{footnotesize}
\begin{verbatim}
> console
[info] Starting scala interpreter...
scala> import miniasm._
scala> Assembler. <tab>
  T        asInstanceOf      initData          isInstanceOf
  main        runPhases         toShort           toString          
  writeContigFile   
scala> val asm = Assembler.initData("-input /export/home/staff/
  gkeeble/temp/ERR015569.1in9.fa -k 31")
scala> ScanReadsPhase(asm)
  Fasta format
  miniasm.genome.bpbuffer.BPKmerSet@7a5cf2b8 Cache hits: 160731328 
  misses: 15668672 ratio: 0.91, rate: 1221.00/ms
scala> BuildGraphPhase(asm)
  15676904 nodes
  15566712 edges
scala> FindPathsPhase(asm)
  ..........
scala> contigs.size
  res7: Int = 5686
scala> contigs.toList.sortWith(_.size > _.size).head.size
  res8: Short = 19032
scala> contigs.toList.sortWith(_.size > _.size).head
  res11: Contig = GGAAGCCACAAAGCCTACATAAATATTCATTCCCTCTGGAGGCA...
\end{verbatim}
\end{footnotesize}

In this interactive session, we first prepare a data object using
\code{Assembler.initData}. This method takes the same parameters that the
Friedrich assembler accepts when run in non-interactive mode. The resulting
\code{AssemblyData} object \code{asm} is then manually passed to different
phases by the user. At any time, the user can construct additional data objects
and compare them or interrogate them more closely. After \code{FindPathsPhase}
has finished running, contiguous paths will be available in the \code{asm} data
object. We can now use the full power of Scala to explore or alter the data
that has become available. First we ask for the number of contigs that were
found (5686).  Then we sort the contigs to have the largest first, defining a
sort function on the fly (\code{\_.size \textgreater \_.size}, called an
\emph{anonymous function}) and asking for its size (19032).  Finally we examine
the actual base pairs in this long contig.

\subsection{Extending the assembler for motif recognition}
\label{sec:extension}
We now show how to extend the assembler with a phase that detects and
displays repeating motifs in the contiguous base pair sequences
(contigs) that have been found. In Scala, \emph{traits} are a basic
unit of composition. Classes and traits can inherit from multiple
traits simultaneously. In this way, one can build up a family of
traits, each one representing a functionality, and compose them as
needed. Phases must extend a basic Friedrich phase trait called
\code{Phase}. In our assembler, which is an application that is
built on top of Friedrich, we define \code{AsmPhase}, which
extends \code{Phase} and make it the convention that all our assembly
phases will extend this new trait. Thus, we now add a phase called
\code{FindRepeatsPhase} (Figure~\ref{fig:findRepeats}).

\begin{figure}
\centering
\lstset{language=scala}
\begin{lstlisting}
class FindRepeatsPhase[T <: Kmer[T]](minTotLen: Int = 8, 
    minMotifLen: Int = 3) extends AsmPhase[T] {
  
    def runImpl(data: AssemblyData[T]): Unit = { 
      println("Finding repeats...")      
      for (c <- data.contigSet.contigs;
          (start, length, pattern) <- c.repeats(minTotLen, 
            minMotifLen)) {
    	 println("Contig: " + c + " pattern: " + 
         pattern + " start offset: " + start)    		  
      } }
}
\end{lstlisting} 
\caption{The newly added FindRepeatsPhase.}
\label{fig:findRepeats}
\end{figure}


The method \code{runImpl} implements the concrete functionality of each phase.
Lines 7-12 show a generalised \emph{for-comprehension}, a special feature of
Scala. The variable \code{c} iterates over all contigs that have previously
been found in the assembly. These are taken from the data object, which has
previously been operated on by other phases. The notation \code{(start, length,
pattern)} declares a 3-tuple of three variables. These will iterate over the
repeated motifs returned by the method \code{c.repeats} for each value of
\code{c}. When repeated motifs are found, they are printed to the console. This
short snippet demonstrates that in many cases, Scala code can be considerably
more compact than corresponding Java code (not shown). Also note the default
values of the parameters: \code{minTotLen: Int = 8, minMotifLen: Int = 3}.
These are the two parameters that constructors for this class take, but since
they have default values, they can be omitted if needed.

Since the \code{runImpl} method can contain any Scala code, it has
access to the full range of Java and Scala APIs. Here we obtain the
desired repeats by using the method \code{c.repeats}, which is defined
in Friedrich's contig class (not shown).  However, we are in no way
limited to using only such built-in methods.

After the new phase has been defined in this fashion, no additional
work is needed. It can be included in pipelines, as shown in
Figure~\ref{fig:pipeline}. It can also be used interactively. We have 
written a small convenience function (not shown) to allow the new phase 
to be invoked by simply referring to its name. After
assembly has been carried out, as shown in our previous interactive
example (Section~\ref{sec:interactive}), we can apply the new phase:

\begin{footnotesize}
\begin{verbatim}
scala> FindRepeatsPhase(res1)
(...) TAGACTTATTAGCGACAATAAAGATTATGAGCCTATCAGTCTGGACGGGGAAGATTT
TGAGATGCTTGGTGTAGTTGTAGGCGAGTTTAAAAGAATGGATTAAAATAGACTTAAGAAAAC
TTTAAGT[TGTCTCCTAGTGTCTCCTAG]TGT...
pattern: TGTCTCCTAGTGTCTCCTAG start offset: 299
\end{verbatim}
\end{footnotesize}


A large number of repeated motifs are found in the contigs that were previously
assembled; we show one of them here. At this point, it is possible to retain
the data that has been produced so far, make adjustments, and assemble again
with different parameters. One can then easily contrast repeated motifs that
are produced by different assembly configurations, all without leaving
Friedrich.

\section{Comparison with other tools and libraries}
\label{sec:comparison}
Friedrich has similarities with many existing tools and frameworks, although we
believe that there are no well-known bioinformatics tools precisely filling
Friedrich's role at the moment. The pipeline and phase structure is similar to
a class of software that might be called \emph{toolkits}. These packages
consist of individual specialised programs that operate on a shared file
format. The user is free to run the programs in any order and can thus create
their own workflow, perhaps through shell scripting. Examples of such toolkits
are SAMtools/Picard~\cite{citeulike:4778506} and GATK~\cite{citeulike:7515828},
for handling nucleotide sequence alignments. While these toolkits come close in
spirit to the Friedrich design, one essential difference is that our ability to
run Friedrich phases in an interactive Scala environment permits users to very
easily inspect and modify data manually in between phases. Unsupported
extensions to a toolkit such as SAMtools require first writing a new program
from scratch, and interactive experimentation would require even more
additional work. Friedrich minimises the cost of free experimentation with data
as it is being processed. Note that Picard and GATK provide Java APIs, which
could be easily integrated into Friedrich.


There are many general frameworks for bioinformatics, such as
BioJava\cite{citeulike:3112014}, BioPython\cite{citeulike:4202607},
BioPerl\cite{citeulike:415502}, BioScala\cite{bioscala} and
BioRuby\cite{citeulike:4461164}. These are all utility libraries of varying
size and scope, aimed at bioinformatics tasks in the respective programming
language. Mangalam provides an informative comparison of the first
three~\cite{citeulike:1022168}. Although this survey is now ten years old, most
of the points it makes about programming language differences are still
essentially valid. However, its conclusion that BioPerl is sufficient for about
90\% of bioinformatics programming needs is now outdated, with the need to
process ever larger data sets. In general, the Bio-* toolkits provide useful
routines and data models but do not prescribe any specific software development
style.  Therefore, they are somewhat orthogonal to our effort, which aims to
provide both an architectural style and foundational libraries to support it.
The Bio-* toolkits can in principle be integrated into Friedrich applications,
in particular BioScala and BioJava. Bioinformatics workflow systems, for
example those provided in Yabi\cite{citeulike:10359904} and
Galaxy\cite{citeulike:7706897}, are user-friendly ways of managing and applying
high level computation pipelines. However, in their focus on ready-made,
finalised modules they are quite different from what Friedrich seeks to become.

\section{Conclusion and remarks}
\label{sec:conclusion}
We have argued for the introduction of a new set of software
development principles for bioinformatics software, and we provide a
framework that supports application development based on these
principles.  We have also shown an existing application based on the
framework.  Principles such as conservation of dimensionality and an
exposed internal structure will allow developers to produce software
that is more useful to bioinformaticians and better suits the research
process. While Friedrich does not aim to provide either the highest
performance or the greatest flexibility, with good levels of both it
represents a new tradeoff that should be considered an important
option for many areas in bioinformatics.

We view the architectural principles presented in
Section~\ref{sec:principles} as essentially complete. However, the
Friedrich software framework is still in an early stage of its
development and many enhancements and extensions have yet to be
implemented. For example, the BioJava\cite{citeulike:3112014}, and
BioScala\cite{bioscala} libraries provide a large amount of
functionality for bioinformatics applications, and when doing so is
suitable, it would be natural to ``wrap'' this functionality as
Friedrich phases, rather than reimplement the functionality from
scratch in Friedrich.



It remains to develop more applications on top of Friedrich,
in addition to the genome assembler we have discussed in this work, in order to
verify that the design principles hold up across a wider range of tasks
in practice.

\bibliography{prib2012}{}

\begin{thebibliography}{10}

\bibitem{citeulike:4202607}
{Cock, P.J.A. et al}.
\newblock {Biopython: freely available Python tools for computational molecular
  biology and bioinformatics}.
\newblock {\em Bioinformatics}, 25(11):1422--1423, June 2009.

\bibitem{citeulike:10005303}
{Compeau, P.E.C., Pevzner, P.A., Tesler, G.}
\newblock {How to apply de Bruijn graphs to genome assembly}.
\newblock {\em Nature Biotechnology}, 29(11):987--991, November 2011.

\bibitem{citeulike:7706897}
{Goecks, J. et al.}
\newblock {Galaxy: a comprehensive approach for supporting accessible,
  reproducible, and transparent computational research in the life sciences.}
\newblock {\em Genome biology}, 11(8):R86+, August 2010.

\bibitem{citeulike:3112014}
{Holland, R.C.G. et al.}
\newblock {BioJava: an Open-Source Framework for Bioinformatics}.
\newblock {\em Bioinformatics}, 24(18):btn397--2097, August 2008.

\bibitem{37122}
{Hundt, R.}
\newblock {Loop Recognition in C++/Java/Go/Scala}.
\newblock In {\em Proceedings of Scala Days 2011}, 2011.

\bibitem{citeulike:10359904}
{Hunter, A. A. et al.}
\newblock {Yabi: An online research environment for grid, high performance and
  cloud computing.}
\newblock {\em Source code for biology and medicine}, 7(1):1+, February 2012.

\bibitem{citeulike:4778506}
{Li, H. et al.}
\newblock {The Sequence Alignment/Map format and SAMtools}.
\newblock {\em Bioinformatics}, 25(16):2078--2079, August 2009.

\bibitem{MacLean2012}
{MacLean, D., Kamoun, S.}
\newblock Big data in small places.
\newblock {\em Nature Biotechnology}, 30(1):33--34, Jan 2012.

\bibitem{citeulike:1022168}
{Mangalam, H.}
\newblock {The Bio* toolkits--a brief overview.}
\newblock {\em Briefings in bioinformatics}, 3(3):296--302, September 2002.

\bibitem{citeulike:7515828}
{McKenna, A. et al.}
\newblock {The Genome Analysis Toolkit: A MapReduce framework for analyzing
  next-generation DNA sequencing data}.
\newblock {\em Genome Research}, 20(9):1297--1303, September 2010.

\bibitem{citeulike:4461164}
{Mitsuteru, N.G. et al.}
\newblock {BioRuby: open-source bioinformatics library}, 2003.

\bibitem{scala}
{Odersky, M.}
\newblock {The Scala Language Specification, Version 2.9}.
\newblock \url{http://www.scala-lang.org/docu/files/ScalaReference.pdf}, May
  2011.

\bibitem{bioscala}
{Prins, P.}
\newblock {BioScala}.
\newblock \url{https://github.com/bioscala/bioscala}, March 2011.

\bibitem{citeulike:9595647}
{Rother, K. et al}.
\newblock {A toolbox for developing bioinformatics software}.
\newblock {\em Briefings in Bioinformatics}, 13(2):244--257, March 2012.

\bibitem{citeulike:4117809}
{Simpson, J.T. et al.}
\newblock {ABySS: a parallel assembler for short read sequence data.}
\newblock {\em Genome research}, 19(6):1117--1123, June 2009.

\bibitem{citeulike:415502}
{Stajich, J.E. et al.}
\newblock {The Bioperl toolkit: Perl modules for the life sciences.}
\newblock {\em Genome research}, 12(10):1611--1618, October 2002.

\bibitem{zerbino2008}
{Zerbino, D.R., Birney, E.}
\newblock {Velvet: Algorithms for de novo short read assembly using de Bruijn
  graphs}.
\newblock {\em Genome Research}, 18(5):821--829, May 2008.

\end{thebibliography}
\bibliographystyle{plain}



\end{document}